# Perpendicular in-plane negative magnetoresistance in ZrTe$_5$

Ning Ma [a, b, 1], Xiao-Bin Qiang [c, 1], Zhijian Xie [a, 1], Yu Zhang [d], Shili Yan [d], Shimin Cao [a, d], Peipei Wang [c], Liyuan Zhang [c], G. D. Gu [e], Qiang Li [e,f], X. C. Xie [a, d], Hai-Zhou Lu [c,] *, Xinjian Wei [a, d,] *, Jian-Hao Chen [a, b, d, g,] *

[a] *International Center for Quantum Materials, School of Physics, Peking University, Beijing 100871, China*
[b] *Key Laboratory for the Physics and Chemistry of Nanodevices, Peking University, Beijing 100871, China*
[c] *Department of Physics and Shenzhen Institute for Quantum Science and Engineering, Southern University of Science and Technology, Shenzhen, China*
[d] *Beijing Academy of Quantum Information Sciences, Beijing 100193, China*
[e] *Condensed Matter Physics & Materials Science Division, Brookhaven National Laboratory, Upton, New York 11973-5000, USA*
[f] *Department of Physics and Astronomy, Stony Brook University, Stony Brook, NY 11794-3800, USA*
[g] *Interdisciplinary Institute of Light-Element Quantum Materials and Research Center for Light-Element Advanced Materials, Peking University, Beijing 100871, China*

[1]*These authors contributed equally to this work.*
*Corresponding Authors: Haizhou Lu(luhz@sustech.edu.cn); Xinjian Wei(weixj@baqis.ac.cn); Jian-Hao Chen(chenjianhao@pku.edu.cn)

## Abstract

The unique band structure in topological materials frequently results in unusual magneto-transport phenomena, one of which is in-plane longitudinal negative magnetoresistance (NMR) with the magnetic field aligned parallel to the electrical current direction. This NMR is widely considered as a hallmark of chiral anomaly in topological materials. Here we report the observation of in-plane NMR in the topological material ZrTe$_5$ when the in-plane magnetic field is both parallel and perpendicular to the current direction, revealing an unusual case of quantum transport beyond the chiral anomaly. We find that a general theoretical model, which considers the combined effect of Berry curvature and orbital moment, can quantitatively explain this in-plane NMR. Our results provide new insights into the understanding of in-plane NMR in topological materials.

ZrTe$_5$ exhibits a variety of topological states, including Dirac semimetal[1], Weyl semimetal[2], strong topological insulator and weak topological insulator[3–5], as well as transitions between these states[3,6–8]. Such extraordinary complexity is due to the fact that the topological properties of ZrTe$_5$ changes with a slight variation in its lattice constant[3], making ZrTe$_5$ an ideal platform for studying the electronic transport behavior of different topological phases and their transitions. In Weyl semimetals, an in-plane magnetic field along the current direction can result in a charge pumping effect between the two different Weyl points, inducing negative magnetoresistance (NMR) proportional to the square of the magnetic field $B^2$, which is referred to as the chiral anomaly[2,9]. In general, such in-plane NMR is regarded as the most significant evidence of chiral anomaly[2,10]. However, besides Weyl/Dirac semimetals, NMR has also been observed in some topological insulators and non-topological materials[11–13], in which chirality may not be well defined. In addition, there are other mechanisms that can also induce NMR, including 1) spin-dependent scattering[14], 2) weak localization[15], 3) current jetting effect[13], 4) transport at the quantum limit[1] and 5) the effect of non-trivial band topology[11–13]. Each of these mechanisms produces different transport characteristics, making it crucial to examine these characteristics when searching for the physical origin of NMR in a specific material.

In this study, we systematically investigate the magnetotransport properties of ZrTe$_5$ thin flakes under an in-plane magnetic field and observe negative magnetoresistance with the magnetic field either parallel or perpendicular to the electrical current direction. Combined with theoretical analysis, we conclude that this previously unreported NMR arises from a combined effect of Berry curvature and orbital moment in ZrTe$_5$.

The ZrTe$_5$ flakes were mechanically exfoliated onto a silicon substrate with a 285 nm oxide layer. Standard e-beam lithography was used to pattern the electrodes, followed by an e-beam evaporation of Cr (5 nm) and Au (60 nm). Electrical measurements were performed using a Physical Property Measurement System (PPMS) with a standard lock-in technique. Careful attention was paid to protect the thin flakes from exposure to ambient conditions. The entire sample preparation and device fabrication processes were carried out in a vacuum, an inert atmosphere, or with the sample capped with a protective layer. This layer consisted of a bilayer of 200 nm polymethyl methacrylate (PMMA) and 200 nm methyl methacrylate (MMA).

We measured several devices that all exhibit similar magneto-transport behavior. In the main text, we focus on a typical one, device 1, which has a thickness of 51 nm. Data from other devices can be found in Supplementary Materials S4. This data shows that the behavior can be well reproduced in devices of different thicknesses. The temperature dependent longitudinal resistance ($R_{xx}$ vs. $T$) of device 1 can be found in Supplementary Materials S1, which exhibits a peak at $T_p$ = 107 K. This resistance peak has been widely observed in ZrTe$_5$ crystals, with $T_p$ ranging from below 2 K to above room temperature[2,6,16,17]. It is associated with temperature-induced Lifshitz transition[7]. The inset of Fig. S1 shows the crystal structure of ZrTe$_5$. The Zirconium atoms run along the *a* axis, and each zirconium atom is related to three tellurium atoms, forming ZrTe$_3$ chains. The ZrTe$_3$ chains are linked by two tellurium atoms along the *c* axis, and the layered structure stacks along the *b* axis by van der Waals interactions. The interlayer van der Waals coupling is weak, thus ZrTe$_5$ can be easily exfoliated into thin flakes[3]. Additionally, since the bonding along the *c* axis is weaker than that along the *a* axis, ZrTe$_5$ thin crystals are typically exfoliated into rectangular shape, with the long edge along the *a* axis[1,2,5,7,16]. In this report, we deliberately fabricate devices with the current direction along the long edge of these rectangular crystals. We defined this direction as *x* axis, and the *z* axis corresponds to the out-of-plane direction (*b* axis).

Longitudinal magnetoresistance of device 1 from 2 K to 300 K in magnetic field ranging from -9 T to 9 T has been investigated along three different field directions. Supplementary Figure S2 shows a large and positive magnetoresistance with the application of a perpendicular magnetic field $B_\perp$, consistent with previous studies[2,6,17]. The magnitude of the positive MR is not always increasing with rising temperature, partly due to the large non-monotonic temperature dependence of the zero-field resistance (Fig. S1). At high temperature (~300 K), the positive MR reduces to about 60% at ~9 T, taking on a form that resembles the conventional positive $B^2$ dependence. This suggests that the high-temperature MR behavior of ZrTe$_5$ thin crystals under $B_\perp$ is likely originated from a classical multi-carrier effect[6]. Figure 1(a) plots MR vs. *B* when the magnetic field is along the *x* axis (marked as $B_{//I}$). In this configuration, NMR is observed, which has previously been interpreted as the signature of chiral anomaly[2]. Surprisingly, when the magnetic field is along the *y* axis (marked as $B_{\perp I}$), a large NMR of more than three times the magnitude of NMR with $B_{//I}$ is observed, as shown in Figure 1(b). The appearance of NMR with $B_{\perp I}$ has not been reported before, and cannot be explained within the theoretical framework of chiral anomaly, which requires the electric and magnetic field to be collinear[2,10]. Furthermore, the large magnitude of NMR as a function of $B_{\perp I}$ excludes the possibility of chiral anomaly resulting from a small misalignment of the magnetic field away from the $y$ axis (hence creating a small $B_{//I}$). Another feature of the NMR vs. $B_{\perp I}$ curves is a kink at $B_{\perp I} \sim 6.5$ T at low temperatures (e.g., $T = 2$ K). This resistance kink is sample-dependent and is not related to the NMR discussed in this paper (see Supplementary Materials S4 for more details).

We compared the NMR with $B_{\perp I}$ and $B_{//I}$ by investigating the temperature-dependent amplitude of the two NMRs at $B_{\perp I} = 9$ T and $B_{//I} = 9$ T, as depicted in Supplementary Figures S3(a) and S3(b), respectively. Similar temperature dependence is found between the two cases, indicating that they have a similar physical origin. Furthermore, the magnetic field angle-dependent MR in both the $x$-$z$ plane and $y$-$z$ plane are also quite similar and are highly sensitive to the angle of the magnetic field, as shown in Supplementary Figures S4.1(a) and S4.1(d), respectively. A slight tilt of the magnetic field away from the sample plane leads to the rapid disappearance of NMR and the appearance of strong positive MR.

In order to better understand the physics behind the observed NMR with $B_{//I}$ and $B_{\perp I}$, we looked into the five possible mechanisms beyond chiral anomaly that can cause NMR as discussed in the introduction section. Firstly, since $ZrTe_5$ is non-magnetic, it is unlikely that spin-dependent scattering is the cause of the NMR observed in our experiment. Indeed, no evidence of the existence of magnetic moment or magnetic order in $ZrTe_5$ has been reported in the extensive literature on this material[16]. Besides, theories of spin-dependent scattering cannot account for the large NMR (up to ~30%) observed in our experiment. Secondly, weak localization[15] is a quantum correction to classical conductance that is sensitive to temperature and can be observed with small magnetic field (typically smaller than 1 T). This mechanism is also unlikely to be the cause of the observed NMR because the signal observed can persist up to ~200 K and at 9 T. Thirdly, NMR can be observed if the current in the crystal is distorted due to strong anisotropy, which usually occurs in 3D crystals where the electrical current is injected through a point contact[13]. However, our samples are no thicker than 100 nm and the flake is fully covered by Cr/Au metal at the source/drain contacts (as shown in the inset of Figure S1), which is unlikely to suffer from current jetting effect[13]. The fourth possibility is that the sample is in the quantum limit, where NMR may be observed[18]. Although no sign of quantum oscillation is observed in our devices, Hall measurements show that the carrier density is of the order of ~$10^{13}$ cm$^{-2}$ (see Supplementary Figures S7(b) and S7(d)). This carrier density is too large for the system to be at the quantum limit at ~4 T, where clear NMR has developed (see Figures 1(a) and 1(b)).

Since the four previously mentioned mechanisms failed to explain the experimental observation, we explore a fifth possibility. For the flake samples used in this experiment, it is nearly impossible to support cyclotron orbits when the magnetic field is applied in-plane. Thus the system can be considered in the semi-classical region, where the motion of electrons can be well described by wave-packet dynamics[19]. As a result, Boltzmann kinetics based on Berry curvature and orbital moment is a suitable method to handle the in-plane magneto-transport behavior in ZrTe$_5$ thin flakes. This approach allows for the examination of nearly all non-quantized transport phenomena, regardless of the direction between the magnetic field and current, even in systems without a clear definition of chirality (e.g., topological insulators). The effect of non-trivial band topology and the influence of Berry curvature on the semi-classical equation of motion for charge carriers can be written as[12,20]:

$$\dot{r} = \frac{1}{\hbar}\nabla_k \tilde{\varepsilon}_k - \dot{k} \times \Omega_k ,  \qquad (1)$$

$$\dot{k} = -\frac{e}{\hbar}(E + \dot{r} \times B) , \qquad (2)$$

where $r$ and $k$ denotes the position and wave vector, $\dot{r}$ and $\dot{k}$ are their time derivatives, $\hbar$ is the reduced Plank constant, $e$ is the elementary charge, $\tilde{\varepsilon}_k = \varepsilon_k - m \cdot B$, $\varepsilon_k$ represents the band dispersion and $m$ is the orbital moment induced by the semi-classical self-rotation of the Bloch wave packet[20], $\Omega_k$ is the Berry curvature. Since the strong Berry curvature is verified by the large anomalous Hall effect (Figs. S7), we assume that it can also have a significant impact on the longitudinal motion of electrons. The in-plane magnetoconductivity derived from the Boltzmann kinetic equation is then given by:

$$\sigma^{xx}_{I//B} = \int \frac{d^3k}{(2\pi)^3} \frac{e^2\tau}{D_k}\left(\tilde{v}^x_k + \frac{e}{\hbar}\tilde{v}^v_k\Omega^v_k B\right)^2 \left(-\frac{\partial \tilde{f}_0}{\partial \tilde{\varepsilon}}\right) , \qquad (3)$$

$$\sigma^{xx}_{I\perp B} = \int \frac{d^3k}{(2\pi)^3} \frac{e^2\tau}{D_k}(\tilde{v}^x_k)^2 \left(-\frac{\partial \tilde{f}_0}{\partial \tilde{\varepsilon}}\right) , \qquad (4)$$

where the dimensionless quantity $D_k = 1 + \frac{e}{\hbar}(\Omega_k \cdot B)$ is the correction to volume of the phase space, $\widetilde{v}_k^x = v_k - \nabla_k(\boldsymbol{m} \cdot \boldsymbol{B})/\hbar$ is the group velocity after the correction of the orbital moment (more details in Supplementary Materials S3).

Figure 2(a) shows the band dispersion of the effective Hamiltonian of ZrTe$_5$ under perpendicular in-plane magnetic field ($\boldsymbol{B} = B_y \hat{\boldsymbol{y}}$ and $B_y = 4\,T$). The effective Hamiltonian is described in Supplementary Materials S8. The corresponding Berry curvature vector field in $k$-space is shown in Figures 2(b)-(d). Note that even when chiral anomaly is presented in the material, the pumped electrons at the Weyl points will still need to rely on their Berry curvature to interact with the external magnetic field and to create an enhancement in the conductivity. Hence our theory is inclusive when considering possible charge pumping effects in the I//B configuration [2,9].

Figure 1(c) and 1(d) present the simulated and experimental NMR curves as a function of $B_{//\mathrm{I}}$ and $B_{\perp\mathrm{I}}$, respectively. The simultaneous good agreement between the experiments and calculations for the two configurations and at four different temperatures indicates that the above semi-classical theory captured the main physics in such anomalous transport phenomena in ZrTe$_5$ thin flakes. Since eqs. (3) and (4) include derivatives of the Fermi-Dirac distribution, calculating the conductivity at low temperatures is computationally demanding. As a result, the calculations in Figure 1 are limited to temperatures between 40 K and 100 K. The simulation parameters used in the calculations can be found in Table S6 of the Supplementary Materials S6. It is worth mentioning that in order to achieve the best fitting results, the chemical potential $\mu$ used in the calculations was slightly different for cases with different in-plane magnetic field directions, despite the same gate voltage and temperature conditions. This difference, which is less than 18 meV, is likely due to the slight deviation of the model Hamiltonian from the real Hamiltonian describing our thin film samples (the model Hamiltonian is taken from bulk ZrTe$_5$, details in Supplementary Materials S8). Considering the delicate correlations between the band topology and Fermi energy with external perturbations such as temperature [7] and strain [8], the slight difference in the Hamiltonian and thus the energy shift in the calculation is within a reasonable range (detailed discussions in Supplementary Materials S8).

In the following we discuss gate-dependent magnetotransport behavior of the ZrTe$_5$ samples. Since the Berry curvature is determined by the position of the Fermi level, it can be expected that the NMR resulting from the Berry curvature can be tuned by the gate voltage. This has indeed been observed in our experiment, as shown in Figure S9. By varying the back gate voltage $V_{bg}$ at $T = 2$ K and $B_{\perp I} = 9$ T, the NMR of device 2 can be electrically tuned by a factor of five, from 5% at $V_{bg}$ = -100 V to 25% at $V_{bg}$ = 100 V. The calculated the dependence of MR on the chemical potential $\mu$ from 40 K to 100 K (Supplementary Figure S6.1) showed that changing the chemical potential can result in substantial changes in the NMR in both B//I and B⊥I configurations at all temperatures. The calculation also showed that the NMR was stronger for the B⊥I configuration compared to the B//I configuration in a lightly hole doping regime. The simulation results showed that the Fermi level of the device was located at the hole side from 40 to 100 K, which is in agreement with the experimentally measured Hall resistance under an out-of-plane magnetic field (as seen in Supplementary Figure S7).

In conclusion, longitudinal negative magnetoresistance is observed in ZrTe$_5$ in both configurations where the in-plane magnetic field is either parallel or perpendicular to the current direction. This magnetoresistance goes beyond the behavior of chiral anomaly. In the recent years, studies of this delicate material ZrTe$_5$ has reported experimental evidence of Dirac semimetal, Weyl semimetal, strong and weak topological insulators[1,4,5]. Among all the studies, one consensus is that ZrTe$_5$ has nontrivial Berry curvature near the Fermi level. Our experiments are well-explained by the combined effects of Berry curvature and orbital moment under the semi-classical equation of motion. Our results provide a unique way to interpret NMR observed in experiments beyond chiral anomaly.

## Conflict of interest

The authors declare that they have no conflict of interest.

## Acknowledgement

The authors thank S. Q. Shen for helpful discussions. This project has been supported by the National Key R&D Program of China (Grant Nos. 2019YFA0308402, 2018YFA0305604), the Innovation Program for Quantum Science and Technology (2021ZD0302403), the National Natural Science Foundation of China (NSFC Grant Nos. 11934001, 92265106, 11774010, 11921005), Beijing Municipal Natural Science Foundation (Grant No. JQ20002). The work at Brookhaven

National Laboratory was supported by the U.S. Department of Energy (DOE), Office of Basic Energy Sciences, Division of Materials Sciences and Engineering, under Contract No. DE-SC0012704.

**Authors contributions**

Jian-Hao Chen designed the overall experiment; Peipei Wang, Liyuan Zhang, G. D. Gu and Qiang Li synthesized and provided the single crystal samples; Zhijian Xie, Xinjian Wei, Yu Zhang, Shili Yan and Shimin Cao carried out the specific experiments; Xiao-Bin Qiang and Hai-Zhou Lu conducted the theoretical analysis and numerical simulations; Ning Ma, Xiao-Bin Qiang, Zhijian Xie and Jian-Hao Chen wrote the manuscript; all the authors discussed the results and comments on the manuscript.

**Figures and Captions**
**Figure 1**

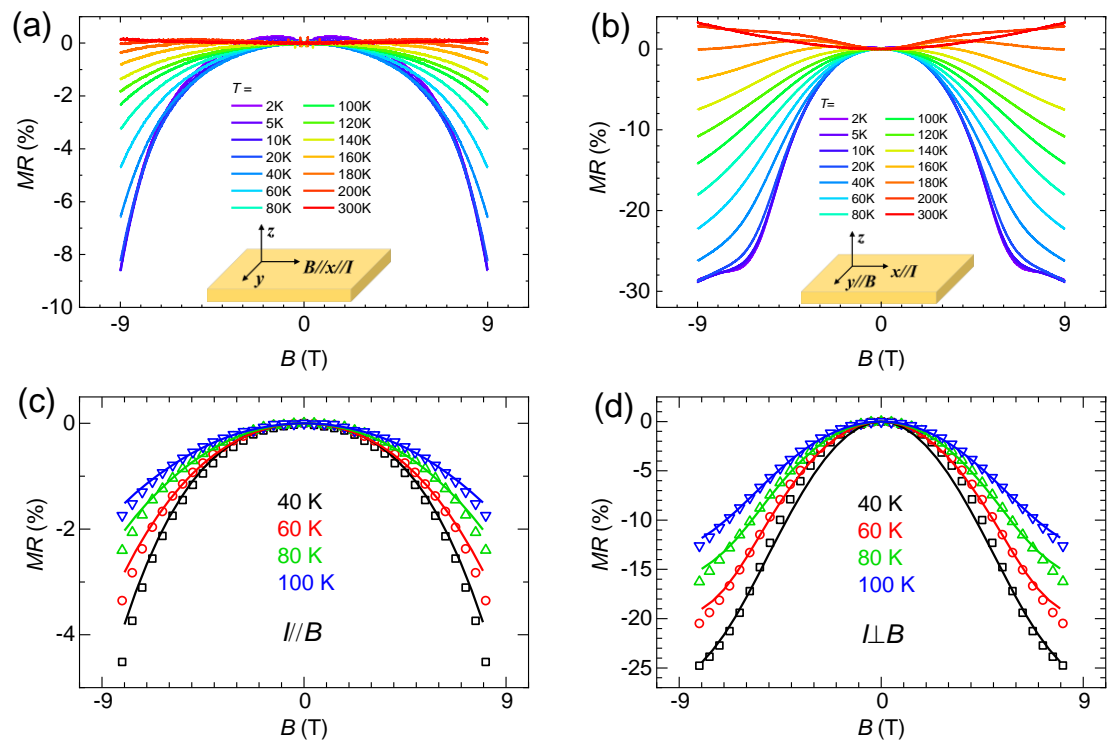

**Figure 2**

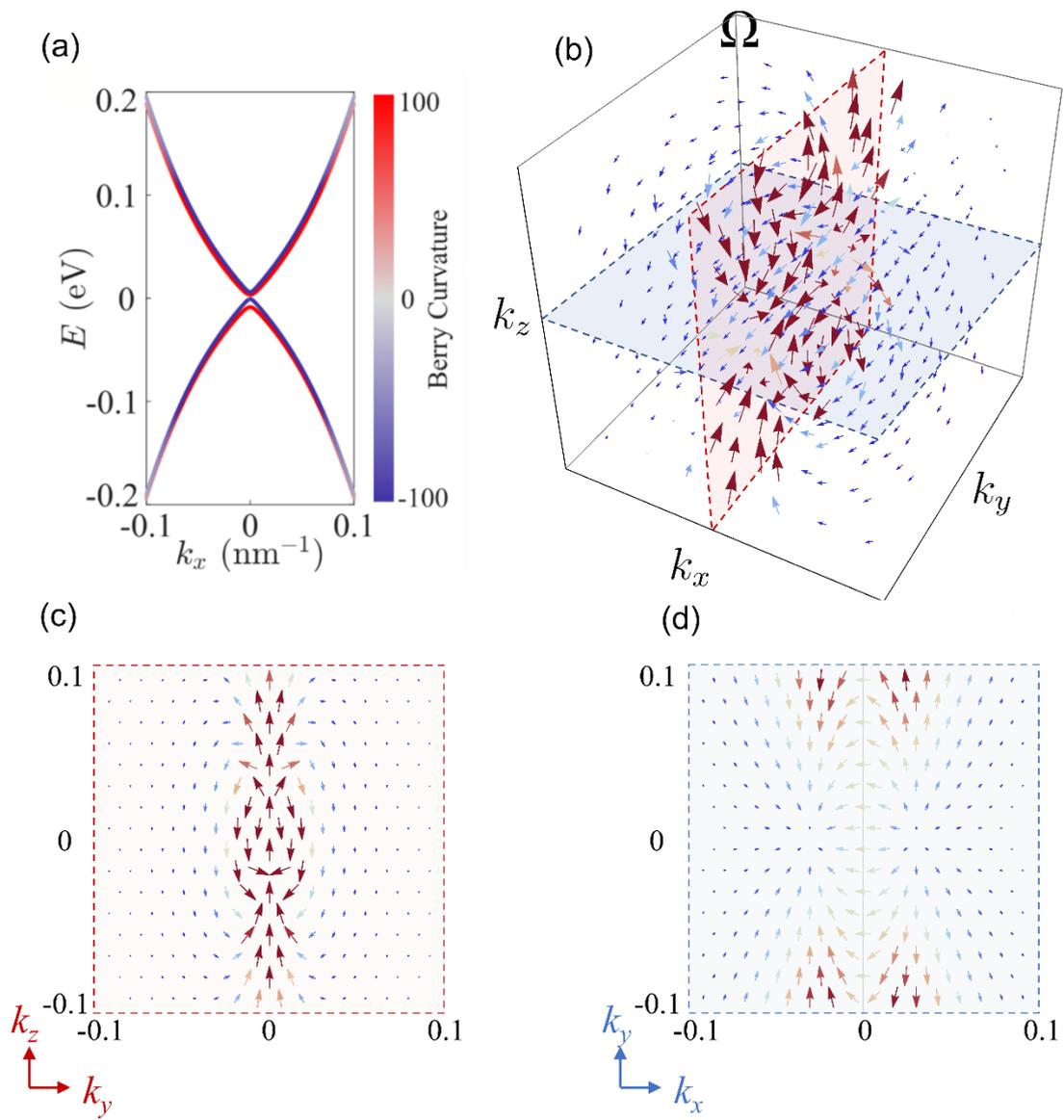

# Figure Captions

**Fig. 1. MR of ZrTe$_5$ at various in-plane magnetic field and the corresponding simulation results.** (a)-(b) Magnetoresistance of device 1 at different temperatures with (a) in-plane magnetic field along the current direction ($\boldsymbol{B}//\hat{\boldsymbol{x}}$), and (b) in-plane magnetic field perpendicular to the current direction ($\boldsymbol{B}//\hat{\boldsymbol{y}}$). The temperature range of (a) and (b) is from 2 K to 300 K. (c)-(d) Simulation (solid lines) and experimental results (symbols) of the negative magnetoresistance of ZrTe$_5$ with in-plane magnetic field (c) parallel and (d) perpendicular to the current direction. The temperature range of (c) and (d) is from 40 K to 100 K.

**Fig. 2. Band dispersion and Berry curvature vector field in ZrTe$_5$.** (a) Band dispersion of the effective Hamiltonian of ZrTe$_5$ under perpendicular in-plane magnetic field ($\boldsymbol{B} = B_y\hat{\boldsymbol{y}}$ and $B_y = 4$ T). The bands splitting is originated from the Zeeman effect, and the color bar on the right depicts the Berry curvature of the energy band. (b) Demonstration of 3D Berry curvature vector field in the *k*-space. (c)&(d) 2D sections of the Berry curvature vector field with $k_x = 0.01\ nm^{-1}$ (c) and $k_z = 0.01\ nm^{-1}$ (d).